\documentclass[aps,prb,twocolumn]{revtex4-1}
\usepackage{amsmath,amssymb,bm}
\usepackage[dvipdfmx]{graphicx}

\usepackage{txfonts}
\usepackage{amsmath}
\usepackage{bm}
\usepackage{comment}
\usepackage{here}
\usepackage{kantlipsum}
\usepackage{color}

\begin{document}

\newcommand{\tm}[1]{\textcolor{black}{#1}}

\title{
Magnetotransport in Layered Dirac 
Fermion System
Coupled with Magnetic Moments
}

\author{
Yoshiki Iwasaki and Takao Morinari}
\affiliation{
Graduate School of Human and Environmental Studies, Kyoto
University, Kyoto 606-8501, Japan}

\date{\today}

\begin{abstract}
We theoretically investigate the magnetotransport of Dirac fermions 
coupled with localized moments to understand the physical properties
of the Dirac material EuMnBi$_{2}$. 
Using an interlayer hopping form,
which simplifies the complicated interaction
between the layers of Dirac fermions and the layers of magnetic moments
in EuMnBi$_{2}$, the theory reproduces most of the features
observed in this system.
The hysteresis observed in EuMnBi$_{2}$ can be caused by
the valley splitting that is induced by
the spin-orbit coupling and the external magnetic field
with the molecular field created by localized moments.
Our theory suggests that the magnetotransport in EuMnBi$_{2}$
is due to the interplay among Dirac fermions,
localized moments, and spin-orbit coupling.
\end{abstract}

\maketitle


Dirac materials have attracted much interest because of their intriguing
topological characteristics.
Unconventional half-integer quantum Hall effect was observed
\cite{Novoselov2005,Zhang2005} in graphene\cite{Geim2007}
due to the Landau level structures of Dirac fermions.
Unlike conventional metals, the backward scattering is strongly
suppressed\cite{Ando1998} due to the Berry phase $\pi$,
which makes Dirac fermions extremely high mobility carriers.


Recently first-principles calculations predicted that
manganese pnictide SrMnBi$_2$ is a Dirac material,
because of the electronic properties
of the Bi square net, and subsequently,
its dispersion and Fermi surface
were observed by
angle-resolved photoemission spectroscopy
and quantum oscillations.\cite{Park2011,Wang2011,Lee2013}
Interestingly, there are structural and physical similarities 
between SrMnBi$_2$ and the 112-type iron-based 
superconductors.\cite{Wang2011}
Antiferromagnetic (AF) ordering of the magnetic moments of Mn
occurs below 290 K as suggested by the temperature dependences
of magnetization, resistivity, 
and specific heat.\cite{Wang2011}
From the perspective of spintronics application,\cite{Zutic2004}
it is important to investigate
the interplay between Dirac fermions and 
magnetic moments.
In this regard, the Dirac material EuMnBi$_2$, 
which is isostructural with SrMnBi$_2$,
provides a plausible platform.\cite{May2014}
Half-integer quantum Hall effect was observed, 
and the Berry phase $\pi$ of Dirac fermions was
found from the analysis of Shubnikov-de-Haas 
oscillations.\cite{Masuda2016}

In EuMnBi$_2$, the layer of Eu$^{2+}$ with spin $S=7/2$ is closer 
to the Bi square net
than the layer of Mn-Bi edge sharing tetrahedra.
AF ordering of Eu moments around $T_{\rm N} = 22$ K 
is suggested from the magnetic susceptibility
measurements.\cite{May2014}
Compared to SrMnBi$_2$,
the N\'{e}el temperature 
associated with the AF ordering of Mn moments
is enhanced to 310 K\cite{May2014},
which is presumably due to the interaction between Mn moments 
and Eu moments.
Transport measurements demonstrated\cite{Masuda2016} that 
the Dirac fermion transport strongly couples to Eu moments.
The ordering of Eu moments is AF 
in the direction of the $c$ axis, which is perpendicular to the layers
of Dirac fermions, and ferromagnetic in the $ab$ plane.
Below 120 K, both the in-plane resistivity $\rho_{xx}$
and the interlayer resistivity $\rho_{zz}$ show metallic behavior
down to $T_{\rm N}$ with the large anisotropy of $\rho_{zz}/\rho_{xx} 
\sim 480$ at 50 K.\cite{Masuda2016}
A small drop in $\rho_{xx}$ and an enhancement in $\rho_{zz}$
were observed at $T=T_{\rm N}$.
The effect of coupling between Dirac fermions and Eu moments
is seen much clearly under a magnetic field.
When the magnetic field is applied in the $c$ axis,
$\rho_{zz}$ increases sharply below $T_{\rm N}$,
and $\rho_{zz}/\rho_{xx}$ exceeds $1\,000\%$ at 9 T,
while it is about 180$\%$ at 0 T.
Spin-flop transition of Eu moments occurs at $\sim $ 5.3 T,
and there is a steep increase in $\rho_{zz}$ at the transition point.
As the magnetic field was increased, 
a peak was observed in $\rho_{zz}$ around 20 T.\cite{Masuda2016}
Remarkably, this peak shows a hysteresis
between the field-increasing and field-decreasing runs.
A hysteretic anomaly was also observed in $\rho_{xx}$.

In this Letter, 
we theoretically study the magnetotransport of two-dimensional
Dirac fermions coupled with localized moments,
and discuss the features experimentally observed in EuMnBi$_{2}$.
We calculate the in-plane and interlayer conductivities by
using the Kubo formula, assuming a phenomenological form of the
interlayer tunneling.
In order to explain the hysteresis observed in EuMnBi$_{2}$,
the lift of valley degeneracy is taken into account.

\begin{figure}[H]
    \begin{center}
     \includegraphics[width=0.8 \linewidth]{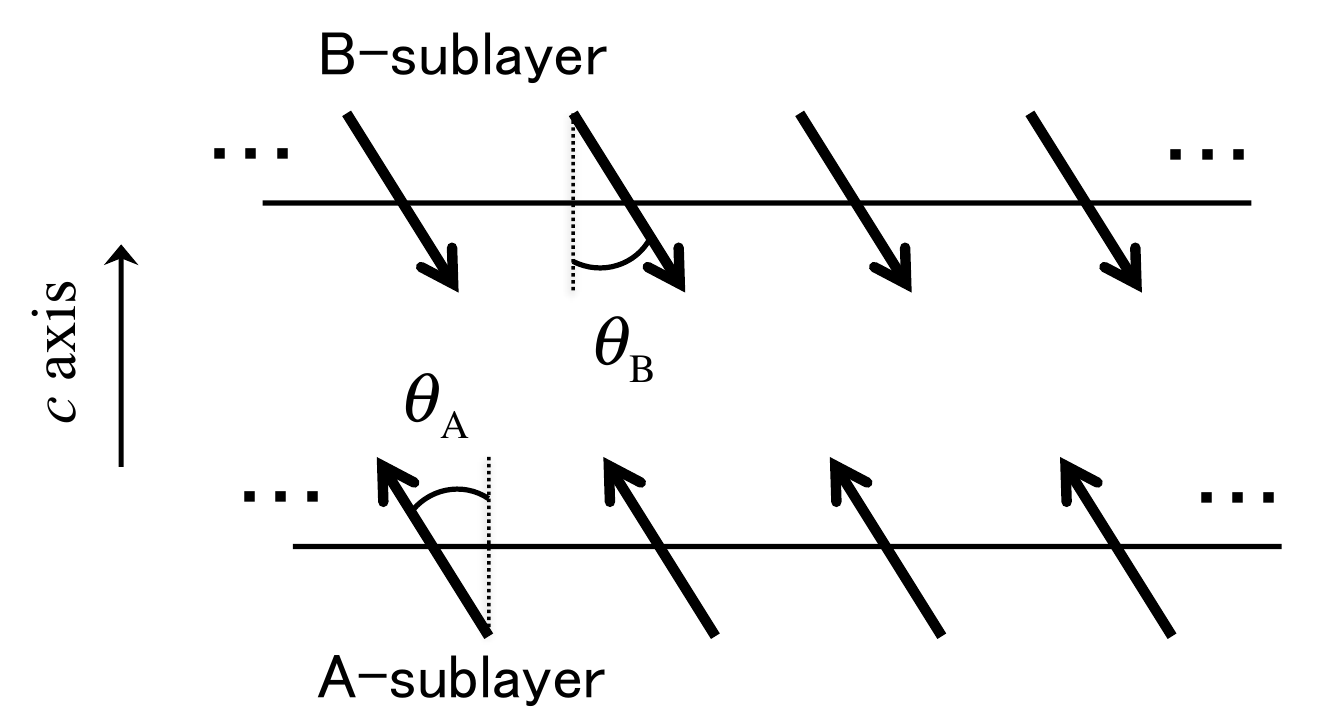}
     \end{center}
\caption{
Layers of localized moments and the definition of the angles,
$\theta_{\rm A}$ and $\theta_{\rm B}$.
In each layer, the localized moments are ferromagnetically ordered, 
while they are antiferromagnetically ordered along the $c$ axis
in the ground state.
In between the layers of localized moments, there is a single
layer of Dirac fermions, which is not shown in the figure.
}
\label{fig:angle}
\end{figure}

\begin{figure}[htbp]
    \begin{center}
\includegraphics[width=0.8 \linewidth]{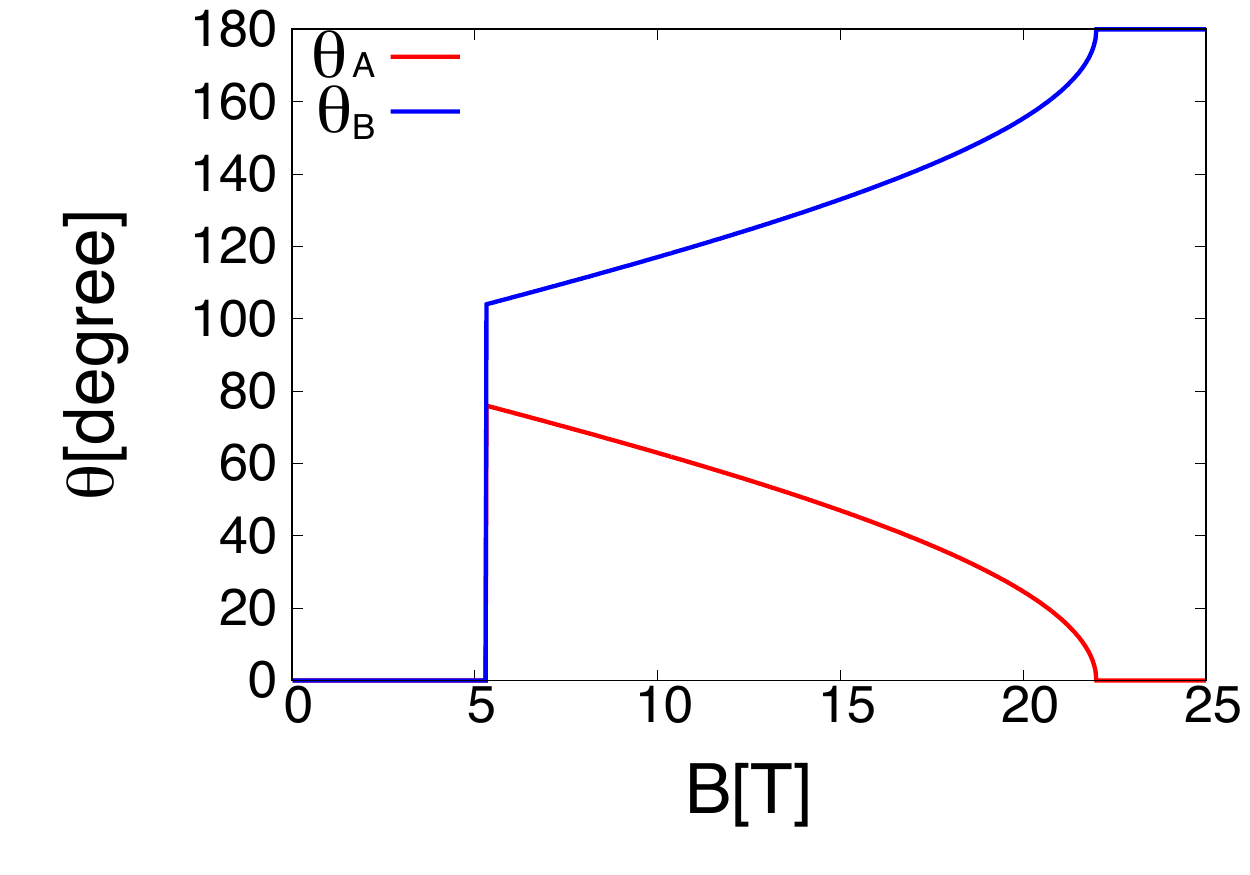}
     \end{center}
\caption{
(Color online)
Magnetic field dependence of the angles of localilzed moments
determined by minimizing the energy (\ref{eq_spin_E}).
Because of the anisotropic energy with the parameter $K$,
there is a spin-flop transition at $B \sim 5$ T.
For $B > 22$ T, the localized moments are fully polarized.
}
\label{fig:theta}
\end{figure}

Before investigating the magnetotransport of Dirac fermions, 
we first consider a model for the layers of magnetic moments
in EuMnBi$_{2}$.
In the absence of a magnetic field,
we assume that the localized moments 
are antiferromagnetically ordered along the $c$ axis
in the ground state.
Therefore, we denote the even layers of localized moments as
A sublayer
and the odd layers of localized moments as B sublayer.
As shown in Fig.~\ref{fig:angle},
we introduce the angles $\theta_{\rm A}$ and 
$\theta_{\rm B}$ to describe
the direction of the localized moments 
in A sublayer and B sublayer, respectively.
The optimum values of $\theta_{\rm A}$ and 
$\theta_{\rm B}$ are determined numerically
by minimizing the following energy:
\begin{eqnarray}
E &=& - K(\cos^{2}\theta_{\rm A}+\cos^{2}\theta_{\rm B})
-\mu_{B}B(\cos\theta_{\rm A}-\cos\theta_{\rm B})\nonumber \\
& & - J\cos(\theta_{\rm A}-\theta_{\rm B}).
\label{eq_spin_E}
\end{eqnarray}
Here, $K$ is the anisotropic energy, $\mu_{B}$ is the Bohr magneton,
$B$ is the magnetic field, and
$J>0$ is the AF interaction between localized moments.
\tm{
The magnetic field dependences of $\theta_{\rm A}$ and $\theta_{\rm B}$
are shown in Fig.~\ref{fig:theta}. 
In the spin-flop phase, we found that the energy is given by
$E =  - J - {\left( {{\mu _B}B} \right)^2}/\left[ {2\left( {J - K}
\right)} \right]$
and 
$\cos {\theta _A} =  - \cos {\theta _B} 
= {\mu _B}B/\left[ {2\left( {J- K} \right)} \right]$.
In terms of the spin flop field $B_f$ and 
the critical field $B_c$, above which the spins are fully polarized,
$K$ and $J$ are given by
$K = {\mu _B}B_f^2/\left( {2{B_c}} \right)$
and
$J = {\mu _B}\left( {B_f^2 + B_c^2} \right)/\left( {2{B_c}} \right)$,
respectively.
Substituting the values $B_f \simeq 5$ T and
$B_c \simeq 22$ T, which were observed 
in the experiment,\cite{Masuda2016}  
we obtain $K=0.43$ K and $J=7.8$ K.
}
In the following, we use this result for the magnetic field dependence
of the directions of localized moments.

For the purpose of calculating 
the in-plane and interlayer conductivities,
we need the Landau level wave functions of Dirac fermions.
For this, we consider a single valley of Dirac fermions.
The presence of another valley and
the effect of the lift of its degeneracy are examined later.
The Hamiltonian is given by
\begin{equation}
 H_D = v\left( {\begin{array}{*{20}{c}}
       0&{{\pi _x} - i{\pi _y}}\\
	       {{\pi _x} + i{\pi _y}}&0
	      \end{array}} \right),
\end{equation}
where $v$ is the Fermi velocity of Dirac fermions and 
${\pi _\alpha } = {p_\alpha } + e{A_\alpha }$
($\alpha = x, y$), with $p_{\alpha}$ and $A_{\alpha}$
being the momentum operators and the vector potential, respectively.
Here, $e$ denotes the electron charge.
\tm{
Based on a first-principles calculation, 
it was pointed out in Ref.~\onlinecite{Park2011}
that the Dirac cone in the Bi square net is anisotropic.
The largest value of the Fermi velocity is
$1.51 \times 10^6 $ m/s
while its smallest value is
$1.91 \times 10^5 $ m/s.
Taking the geometrical mean of these values,
we assume, $v=5\times$10$^{5}$ m/s.
}

Taking the Landau gauge, $\bm{A}=(0,Bx,0)$,
we assume a plane wave function in the $y$-direction.
The energy of the Landau levels is given by\cite{CastroNeto2009}
\begin{equation}
E_{n}=\mbox{sgn}(n)v\sqrt{2e\hbar B \lvert n \rvert},
\end{equation}
with integer $n$.
The wave function of the Landau level with $n$ 
is denoted by $\bm{F}_n(\bm{r})$, where
\begin{equation}
 \bm{F}_{0}(\bm{r})=\left(
		\begin{array}{c}
		 0 \\
		 h_{0}(\bm{r}) \\
		\end{array}
  \right)
 \label{eqF0}
\end{equation}
and
\begin{equation}
 \bm{F}_{n}(\bm{r})=\frac{1}{\sqrt{2}}\left(
    \begin{array}{c}
      \mbox{sgn}(n) h_{\lvert n\rvert-1}(\bm{r}) \\
      h_{\lvert n\rvert}(\bm{r}) \\
    \end{array}
  \right)
 \label{eqFn}
\end{equation}
for $n\neq0$.
The function $h_{|n|}(\bm{r})$ is given by
\begin{eqnarray*}
h_{\lvert n\rvert}(\bm{r})
&=& i^{n}\sqrt{\frac{1}{\sqrt{\pi}2^{\lvert n\rvert}\lvert n\rvert !
\ell L}}
H_{\lvert n\rvert}\left(\frac{x-k\ell^2}{\ell}\right)\\
& & \times
 \exp\left[-\frac{1}{2}\left(\frac{x-k\ell^2}{\ell}\right)^{2}\right]
 \exp(iky).
\end{eqnarray*}
Here, $H_{\lvert n\rvert}(x)$ is the Hermite polynomial, 
$\ell=\sqrt{\hbar/e|B|}$ is the magnetic length,
and $L$ is the system size in the $y$ direction.

The interlayer conductivity is computed 
by the Kubo formula.\cite{Osada2008}
The result is,
\begin{eqnarray}
\sigma_{zz}=\frac{2t_c^{2} a_c e^{3}\tau}{\pi\hbar^{3}}B\rho(\mu).
\label{eq_sigma_zz}
\end{eqnarray}
Here, $t_{c}$ is the interlayer hopping, 
$a_c = 22.5\AA$ is the lattice constant in the $c$
axis\cite{Masuda2016},
$\tau$ is a constant relaxation time, 
$\mu$ is the chemical potential,
and $\rho(\epsilon)$ is the spectral function, which has
the form of a Lorentz function.
\tm{
From the analysis of the temperature dependence 
of the Shubnikov-de Haas oscillation amplitude
in $\rho_{xx}$, $\tau$ was estimated 
as $\tau = 3.5\times10^{-14}$ s
in SrMnBi$_2$.\cite{Park2011}
We found that this value is too small for EuMnBi$_{2}$,
and therefore we assumed that $\tau = 2.5\times10^{-13}$ s
to reproduce the Shubnikov-de Haas oscillation 
observed in Ref.~\onlinecite{Masuda2016}.
}
\tm{
For the value of $\mu$, we assume $\mu=1\,000$ K
to reproduce the Shubnikov-de Haas oscillation period that was
observed experimentally in $\rho_{xx}$ and $\rho_{zz}$.\cite{Masuda2016}.
}

In order to apply the formula (\ref{eq_sigma_zz})
to EuMnBi$_{2}$, we need to determine
the interlayer hopping parameter, $t_c$.
In between the Dirac fermion layers of EuMnBi$_{2}$,
there are three layers: 
two layers of Eu moments
and one layer composed of Mn$^{2+}$ ions and Bi$^{3-}$ ions.
The parameter $t_c$ depends on not only 
the directions of Eu moments and Mn moments
but also the interaction between Dirac fermions and Bi$^{3-}$ ions.
We note that those Bi$^{3-}$ ions are different 
from Bi$^{1-}$ ions forming
the layers of Dirac fermions.\cite{May2014,Masuda2016}
In order to avoid this complication, here we 
assume that the magnetic layers between Dirac fermion layers
consist of a single component of localized moments,
whose AF order is described by $\theta_{\rm A}$ and 
$\theta_{\rm B}$, above.
\tm{
Then, in the spin-flop phase,
$t_{c}$ depends on $\theta_{\rm B} = \pi - \theta_{\rm A}$.
}
Apparently, the interlayer hopping $t_{c}$ 
is maximum when the localized moments are fully polarized,
while it is minimum in the AF phase. 
As a simple phenomenological model, 
we assume the following form
\begin{equation}
t_{c}\propto 1-\cos\frac{\theta_{\rm B}}{2}.
\label{eq_tc}
\end{equation}
The proportionality constant is chosen such that
$t_{c}$ takes the value of 200 K 
in the fully polarized state of localized moments.

The Kubo formulae for the in-plane conductivity 
and Hall conductivity are given by
\begin{equation}
\sigma_{xx}=\frac{e^{2}\hbar}{2 \ell^{2}a}\sum_{n,n'}
 \rho(\mu-E_{n})\rho(\mu-E_{n'})\lvert\langle \bm{F}_{n}\lvert
 \sigma_{x}\rvert \bm{F}_{n'}\rangle\rvert^{2},
\end{equation}
\begin{equation}
\sigma_{xy}=\frac{\hbar e^{2}v^{2}}{\pi a \ell^{2}}\sum_{n,n'}
 \frac{f(\mu-E_{n})}{(E_{n}-E_{n'})^{2}}{\rm
 Im}\langle\bm{F}_{n}\lvert \sigma_{x}\rvert
 \bm{F}_{n'}\rangle\langle\bm{F}_{n'}\lvert
 \sigma_{y}\rvert \bm{F}_{n}\rangle,
\end{equation}
respectively.
Here, $a=4.5\AA$ is the lattice constant in the plane,\cite{May2014}
and $\sigma_{x}$ and $\sigma_{y}$ are Pauli matrices.
The function $f(x)$ is defined by
$f(x)=1/2+\arctan\left(\frac{x}{\Gamma}\right)/\pi$,
where
$\Gamma=\hbar/\tau$ is 
a constant parameter describing the broadening
of Landau levels due to impurity scatterings.
The resistivity is computed numerically using the formula given by
$\rho_{xx}=\sigma_{xx}/(\sigma_{xx}^{2}+\sigma_{xy}^{2})$,
${\rho _{yx}} 
= {\sigma _{xy}}/\left( {\sigma _{xx}^2 + \sigma _{xy}^2} \right)$,
and 
$\rho_{zz}=1/(\sigma_{zz}+\sigma_{0})$.
\tm{
Here, parameter $\sigma_{0}=0.01$,
which is associated with impurity scattering,
is introduced to reproduce the experimentally 
observed $\rho_{zz}$.\cite{Masuda2016}
}

\begin{figure}[htbp]
    \begin{center}
\includegraphics[width=0.8 \linewidth]{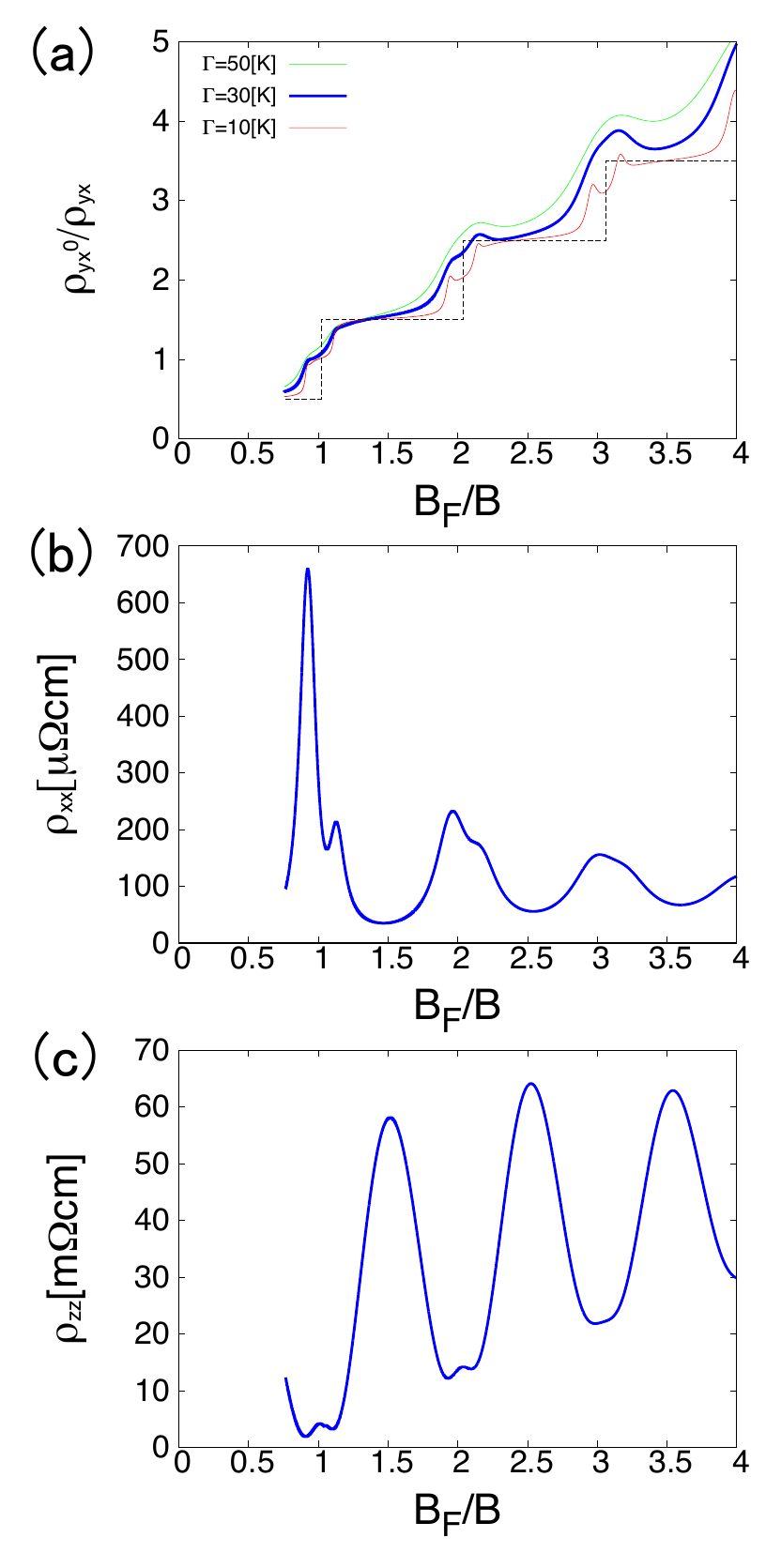}
     \end{center}
\caption{
(Color online)
Normalized inverse Hall resistivity $\rho_{yx}^0\rho_{yx}/$ (a),
in-plane resistivity $\rho_{xx}$ (b),
and interlayer resistivity $\rho_{zz}$ (c)
versus the inverse of the normalized magnetic field.
Here, we take $\rho_{yx}^0 = 1.1\times 10^{-3}$ $\Omega \cdot $cm.
$B_F$ is the frequency of Shubnikov-de-Haas oscillations
defined in the main text.
\tm{
Here, we plot the values in $B<B_c$ where $t_c$, 
which is given by Eq.~(\ref{eq_tc}), is finite.
}
\tm{
In the panel (a) we plot $\rho_{yx}$ for different values of $\Gamma$
and the idealized case is shown in the dashed line.
}
}
\label{fig:resistivity}
\end{figure}

The result obtained by
including the Zeeman energy splitting
is shown in Fig.~\ref{fig:resistivity}.
Here, the magnetic field is normalized by 
introducing parameter 
${B_F} \equiv {\mu ^2}/\left( {2e\hbar {v^2}} \right)
\simeq 23$ T,
which is the frequency of 
Shubnikov-de-Haas oscillations.
The inverse of the Hall resistivity $\rho_{yx}$
shows a half-integer quantum Hall effect,
as shown in Fig.~\ref{fig:resistivity}(a).
Some deviations from the ideal quantum Hall plateaus
occur due to the broadening factor, $\Gamma$,
and similar features were seen in the experiment.\cite{Masuda2016}
$\rho_{xx}$ and $\rho_{zz}$ are shown
in Fig.~\ref{fig:resistivity}(b) and (c),
respectively.
\tm{
The positions of the peaks of $\rho_{xx}$ and the minimum of $\rho_{zz}$
reflect the Landau level structure.\cite{Masuda2016}
}
The density of states takes a large value
when the chemical potential is equal to 
a Landau level energy upon varying the magnetic field.
Reflecting this, $\rho_{zz}$ shows a dip
while $\rho_{xx}$ shows a peak
around integer values of $B/B_F$.
The splitting of peaks in $\rho_{xx}$
and that of dips in $\rho_{zz}$ 
are due to the lift of the level degeneracy.
Here, the degeneracy is associated with the spin degrees of freedom.
The result shown in Fig.~\ref{fig:resistivity}
qualitatively agrees with the experiment.\cite{Masuda2016}
In the experiment, the splitting of the peak in $\rho_{xx}$ 
is observed at $B_F/B=1$.
Furthermore, the peak at $B_F/B<1$
is larger than that at $B_F/B>1$.
This is consistent with the experiment as well.
In the experiment,
the splitting of the peak in $\rho_{xx}$ 
is not observed
at $B_F/B=2$.
However, the second derivative of $\rho_{xx}$ 
with the minus sign, obtained
from the experimental data, shows the splitting
of the peak.
\tm{
The positions of the minimum in $\rho_{zz}$, shown 
in Fig.~\ref{fig:resistivity}(c),
is also consistent with the experiment.\cite{Masuda2016}
}
However, there is an important difference:
the theory suggests the presence of two dips 
and one small peak around $B_F/B=1$,
whereas this peak is much larger 
in the experiment.
It is unlikely that the peak appears
as a result of splitting of the dip.

\begin{figure}[htbp]
    \begin{center}
\includegraphics[width=0.8 \linewidth]{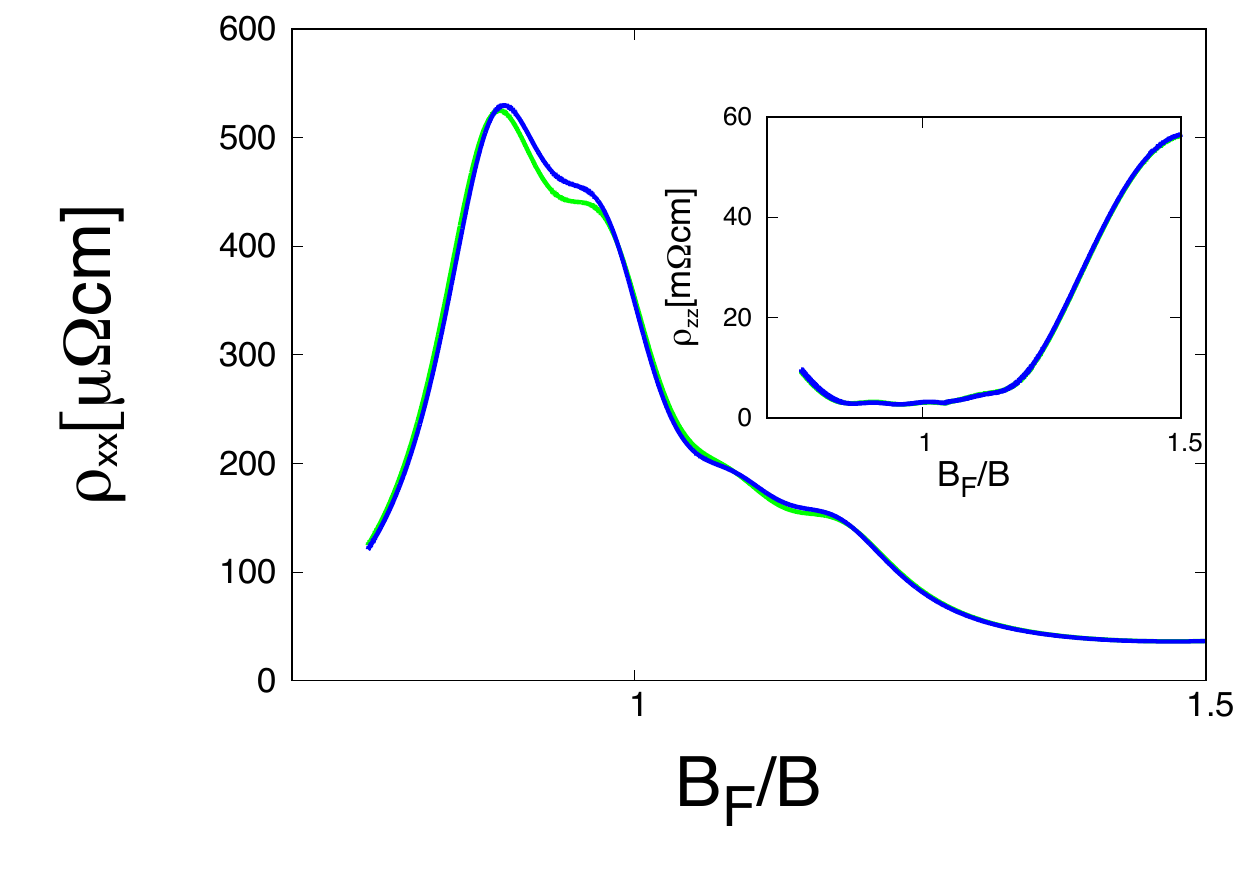}
     \end{center}
\caption{
(Color online)
In-plane resistivity $\rho_{xx}$
and interlayer resistivity $\rho_{zz}$ (inset)
versus the inverse of the normalized magnetic field.
The green lines indicate the result with $\Delta = 1.6\mu_B$,
and the blue lines indicate the result with $\Delta = 1.5\mu_B$.
}
\label{Fig:valley}
\end{figure}

Now we consider the effect of valley splitting.
The Hamiltonian is given by
\begin{equation}
H = v\left( {\begin{array}{*{20}{c}}
{\tilde \Delta }&{{\pi _x} - i{\pi _y}}&0&0\\
{{\pi _x} + i{\pi _y}}&{\tilde \Delta }&0&0\\
0&0&{ - \tilde \Delta }&{{\pi _x} + i{\pi _y}}\\
0&0&{{\pi _x} - i{\pi _y}}&{ - \tilde \Delta }
\end{array}} \right).
\label{eq:valley}
\end{equation}
Here, $\tilde{\Delta}=\Delta/v$.
The parameter $\Delta$ is the energy gap
created by the valley splitting.
The Landau level wave functions for another valley 
is obtained by simply multiplying $\tau_x$,
which is the Pauli matrix
in the sublattice space in the layer of Dirac fermions,
to the two-component spinors,
Eqs.~(\ref{eqF0}) and (\ref{eqFn}),
from the left hand side.
Experimentally, it is suggested that $\Delta$ 
depends linearly on the magnetic field.\cite{MasudaPrivate}
We assume that the valley splitting occurs
through the interaction with the localized moments,
and the origin of the hysteresis 
is in the system of localized moments.
Therefore, we use different proportionality constants
for the field-decreasing run and the field-increasing run.
We use $1.6\mu_B$ for the former and $1.5\mu_B$ for the latter.
\tm{
We used these values to reproduce
the two-peak structure of $\rho_{xx}$ 
that was observed
around $B_F/B = 1$ 
in the experiment.\cite{Masuda2016}
The difference of $0.1 \mu_B$ in these values is taken 
such that
the numerical calculation reproduces 
the experimentally observed difference in peak values,
which are associated with the hysteresis,
of $\rho_{xx}$ at $B_F/B \sim 1.2$.
}
The numerical calculation result is shown in Fig.~\ref{Fig:valley}.
The difference between the two cases
is seen in $\rho_{xx}$ around $B_F/B \sim 1$,
which is similar to that observed in the experiment.\cite{Masuda2016}
However, the theory failed to reproduce the experimentally observed
feature in $\rho_{zz}$ around $B_F/B \sim 1$.\cite{Masuda2016}
This discrepancy is probably due to the form
of the interlayer hopping Eq.~(\ref{eq_tc}).
We need a more realistic $t_c$ than Eq.~(\ref{eq_tc})
to explain the experiment,
which is left for future research.

An important question,
which needs to be considered, 
is the mechanism of valley splitting.
If we recall that bismuth plays a major role
in topological insulators\cite{Hasan2010,Qi2011}
and the key is strong spin-orbit coupling,
one possible scenario is that valley splitting occurs 
through spin-orbit coupling, 
$i{\boldsymbol \lambda}  \cdot {\boldsymbol \sigma}$
with 
${\boldsymbol \lambda}=(\lambda_x,\lambda_y,\lambda_z)$.
Generally, ${\boldsymbol \lambda}$ depends on the wave vector
of Dirac fermions; however,
if we focus on a single Dirac point, then the Hamiltonian is
given by
\begin{equation}
{H_\lambda } = \left( {\begin{array}{*{20}{c}}
{ - {\mu _B}{B_{{\rm{eff}}}}}&0&{i{\lambda _z}}&{i{\lambda _x} +
 {\lambda _y}}\\
0&{{\mu _B}{B_{{\rm{eff}}}}}&{i{\lambda _x} - {\lambda _y}}&{ -
 i{\lambda _z}}\\
{ - i{\lambda _z}}&{ - i{\lambda _x} - {\lambda _y}}&{ - {\mu
 _B}{B_{{\rm{eff}}}}}&0\\
{ - i{\lambda _x} + {\lambda _y}}&{i{\lambda _z}}&0&{{\mu
 _B}{B_{{\rm{eff}}}}}
\end{array}} \right).
\end{equation}
Here, we have included in $B$, the effect of the molecular fields
created by localized spins,
and we denote the effective magnetic field by $B_{\rm eff}$.
The eigenvalues of this Hamiltonian is obtained exactly as,
\tm{
$E^{(\pm,\pm)} = 
\pm \sqrt {{{\left( {{\lambda _z} \pm 
{\mu _B}{B_{{\rm{eff}}}}}\right)}^2} + \lambda _x^2 + \lambda _y^2}$.
}
For $|\lambda_z| \gg |\mu_B B_{\rm eff}|$,
we obtain,
\tm{
 ${E^{\left( { \pm , \pm } \right)}} \simeq  \pm \lambda  \pm
  \frac{{{\lambda _z}}}{\lambda }{\mu _B}{B_{{\rm{eff}}}}$.
}
Therefore, in this scenario, the parameter $\Delta$ is given by
\begin{equation}
 \Delta  = \frac{{2{\lambda _z}}}{\lambda }{\mu _B}{B_{{\rm{eff}}}}.
\end{equation}
If $2{\lambda _z}/\lambda \sim 1$, then the above calculation
is justified.
Furthermore, the slight change in $\Delta$ in
the field-decreasing run and the field-increasing run
is associated with the hysteresis in the system
of localized moments.

To conclude, we have investigated 
the magnetotransport of two-dimensional
Dirac fermions coupled with localized moments,
and compared the theoretical result with the experimental result
for EuMnBi$_{2}$.
Most of the features observed in EuMnBi$_{2}$ 
is understood by our model with interlayer
hopping Eq.~(\ref{eq_tc}).
The hysteresis observed in $\rho_{xx}$ and $\rho_{zz}$
can be associated with the valley splitting
resulting from the spin-orbit coupling
and the coupling between Dirac fermions and localized moments.
However, the theory failed to explain the feature 
around the Landau level with the index $n=1$.
Presumably, this discrepancy arises from
the complicated interaction between
Dirac fermions and the magnetic layer
in EuMnBi$_{2}$.

\section*{Acknowledgments}
We thank H. Masuda and S. Sato for helpful discussions.
This work was supported 
 by a Grant-in-Aid for Scientific Research 
(B) (No. 25287089), 
from the Ministry of Education, 
Culture, Sports, Science, and Technology, Japan.
\bibliographystyle{apsrev4-1}

\end{document}